\documentclass[twocolumn,showpacs,preprintnumbers,amsmath,amssymb,superscriptaddress]{revtex4}

\usepackage{graphicx}
\usepackage{dcolumn}
\usepackage{bm}

\begin{document}


\title{Temperature and Magnetic Field Dependencies\\in a Model of Diamagnetic Hysteresis in Beryllium}

\author{Nathan Logoboy}

\email{logoboy@phys.huji.ac.il}

\affiliation{Grenoble High Magnetic Field Laboratory, MPI-FKF and
CNRS P.O. 166X, F-38042 Grenoble Cedex 9, France}

\affiliation {The Racah Institute of Physics, The Hebrew University
of Jerusalem, 91904 Jerusalem, Israel}

\author{Walter Joss}
\affiliation{Grenoble High Magnetic Field Laboratory, MPI-FKF and
CNRS P.O. 166X, F-38042 Grenoble Cedex 9, France}

\affiliation {Universit$\acute{e}$ Joseph Fourier, B.P. 53, F-38041
Grenoble Cedex 9, France}

\date{\today}

\begin{abstract}
The model of diamagnetic hysteresis loop of strongly correlated
electron gas at the conditions of dHvA effect is developed. It is
shown, that in the framework of Rayleigh theory for magnetization
loop, the coercive force and remnant magnetization in every period
of dHvA oscillations are characterized by strong dependencies on
temperature, magnetic field and Dingle temperature.
\end{abstract}

\pacs{75.20.En, 75.60.Ch, 71.10.Ca, 71.70.Di, 71.25.-s; 71.25. Hc; 75.40-s; 75.40.Cx.}
\maketitle

\section{\label{sec:Introduction}Introduction}

At high magnetic field and low temperature the strongly correlated
electron gas is characterized by instability due to magnetic
interaction between electrons \cite{Shoenberg}, which results in
phase transition from homogeneous state into inhomogeneous one and
stratification of the sample into diamagnetic domains, e.g. Condon
domains (CDs), with alternating from one domain to another
diamagnetic moment. This instability of an electron gas is called
diamagnetic phase transition and extensively studied, both
theoretically and experimentally \cite{Condon}-\cite{Solt2}. The
diamagnetic domains were observed in Be by magnetization
measurements \cite{Condon}; in Ag by nuclear magnetic resonance
(NMR) \cite{Condon_Walstedt} and recently by Hall probe spectroscopy
\cite{Kramer1}. They have been observed in Be, Sn, Al and Pb by muon
spin-rotation ($\mu$SR) spectroscopy \cite{Solt1}. Usually, for
plate-like samples, this non-uniform phase is realized as periodic
domain structure \cite{Shoenberg}-\cite{Kramer1} with alternating
from one domain to another diamagnetic moment. Magnetic interaction
of strongly correlated electrons gas results in non-linear
relationship between local diamagnetic moments and external magnetic
field. It was shown \cite{Logoboy1} that in CD phase the local
diamagnetic moments, which contribute to the process of
magnetization, are characterized by strong temperature and magnetic
field dependencies.

The existence of metastable states for local diamagnetic moments in
each period of dHvA oscillations can results in hysteresis, e. g.
irreversible changes in magnetization due to change in external
magnetic field, similarly to magnetic materials of spin origin. The
microscopic mechanism of magnetization loop, relates to defects of
crystal lattice, intrinsic stress and elastic deformations. The
impurities, dislocations and other defects of crystal lattice cause
an internal friction which opposes to the wall displacement. The
domain walls are pinned by such defects. Unpinning is realized by
small dissipative jumps through the increasing of the value of
applied magnetic field.

Although, the possibilities of irreversible behavior of
magnetization curve in diamagnetic materials have been discussed
earlier \cite{Shoenberg}, only recent experimental investigation of
CDs in Be \cite{Kramer2} reveals hysteresis loop in the
magnetization process. The hysteresis magnetization curve of Be
\cite{Kramer2} was measured directly by Hall probes in dc field and
standard ac method with different modulation levels, frequencies and
magnetic field ramp rates, and the shape of magnetization curve was
reconstructed by several higher harmonics of the ac pickup voltage.
Detection of the hysteresis in Be \cite{Kramer2} substantiates a far
reaching analogy between an electron gas, which has unstable,
metastable and stable states within one dHvA period, and the
liquid-gas phase coexistence and transformation. Recent progress
\cite{Kramer1}, \cite{Kramer2}  in experiment on observation of
Condon domain structure in Ag and Be, in particular, the detection
of diamagnetic hysteresis, provides a natural stimulus towards a
more detailed understanding of the properties of strongly correlated
electron gas in the conditions of dHvA effect. CDs in a $3d$
electron gas are the only type of magnetic structures, for which the
magnetization process has not been considered so far.

In present paper we develop the model of diamagnetic hysteresis,
which fits the recent data on observation of diamagnetic hysteresis
in Be \cite{Kramer2}. The hysteresis loop appears periodically in
every period of dHvA oscillations, when $\mu_{0}\chi
>1$ ( $\chi$ is differential magnetic susceptibility, $M$ is the oscillatory
part of the magnetization, and $B$ is magnetic induction
\cite{Shoenberg}). There are three main contributions into the
process of magnetization of normal metals in the condition of strong
dHvA effect which have to be distinguished, the dependence of local
diamagnetic moments on external magnetic field through the reduced
amplitude of dHvA oscillations \cite{Logoboy1}, the domain wall
displacement and nucleation of the domains. The rotation processes,
that are intrinsic property of magnetization loop in magnetic
materials of spin origin, can be neglected in rearrangement of CD
structure at least for metals with relatively simple Fermi surfaces
\cite{Shoenberg} due to one-dimensional character of the diamagnetic
moments. We show that the remnant magnetization and coercivity, as
well as the hysteresis loop square, which are the fundamental
properties of the magnetization curve, have strong dependence on
temperature, magnetic field and Dingle temperature.

The paper is organized as follows. Section \ref{sec:Introduction} is
an introduction. In Section \ref{sec:Model} we consider the model of
diamagnetic hysteresis, based on Rayleigh model in the framework of
Lifshitz-Kosevich-Shoenberg approximation. In Section
\ref{sec:Results and Discussions} we calculate the temperature and
magnetic field dependence of remnant magnetization and coercivity.
Section IV is conclusions. The last section contains
acknowledgements.

\section{\label{sec:Model}Model}

The oscillator part of the thermodynamic potential density in the
Lifschitz-Kosevich-Shoenberg formalism \cite{Shoenberg} can be
written in reduced form with taking into account the shape sample
effects:

\begin{equation} \label{eq:free energy density}
\Omega =\frac{1} {4 \pi k^{2}} \left [ a \cos {b}+\frac{1}{2}a^{2}(1-n) \sin^{2} {b} \right ], \\
\end{equation}
where $b=k(B-\mu_{0}H)=k[\mu_{0}h_{ex}+(1-n)M]=x+(1-n)y$, $\mu_{0}H$
is the magnetic field inside the material at the center of dHvA
period, $k=2 \pi F/(\mu_{0}H)^{2}$, $F$ is the fundamental
oscillation frequency), $H_{ex}$ is external magnetic field,
$h_{ex}=H_{ex}-H$ is the small increment to the magnetic field $H$,
$n$ is the demagnetization factor. All the components of vectors are
taken along the direction of the magnetic induction. The quality
$a=4 \pi k A=\mu _{0} \partial M/\partial B$ \cite{Shoenberg} is the
reduced amplitude of the dHvA oscillations (the ratio between the
amplitude of oscillations $A$ and their period $\Delta H$).

In the first harmonic approximation the magnetization is found from
the implicit equation of state \cite{Shoenberg}:

\begin{equation} \label{eq:magnetization}
y=a \sin {[x+(1-n)y]}, \\
\end{equation}
 The validity of thermodynamic potential density $\Omega$ (\ref{eq:free energy density}), and the expression for the reduced
magnetization $y$ (\ref{eq:magnetization}), derived from
Eq.~(\ref{eq:free energy density}), is restricted by applicability
to homogeneous phase only, where the existence of demagnetization
coefficient $n$ is justified. In the conditions of strong magnetic
interaction between electrons, when

\begin{equation} \label{eq:CDCondition}
a \left ( \mu_{0}H, T, T_{D}\right )\ge 1, \\
\end{equation}
a state of lower thermodynamic potential is achieved over part of
dHvA oscillation cycle by the formation of CD structure, for which
the local value of magnetization alternates in sign from one domain
to the next, resulting in essential reduce of magneto-static energy.
The usually observed diamagnetic domain structure is of
stripe-domain type \cite{Shoenberg},\cite{Kramer1}, and the
contribution of magneto-static energy into the free energy density
is negligibly small in comparing to the magneto-static energy of
homogeneous state. In the domain state, when the reduced amplitude
of dHvA oscillations satisfies to the condition
(\ref{eq:CDCondition}), the demagnetization factor $n$, which is
characteristic of the uniformly magnetized sample, is replaced by
the coefficient $\alpha$ \cite{Logoboy1}. This coefficient depends
on magnetic field, temperature, impurities and geometry of the
sample and takes into account the magneto-static energy for given
type of domain structure. Typically for plate-like sample,
$\alpha<<1$, independently on the types of domain structures
\cite{Kittel} and important mainly in calculation of the
steady-state period $2D$ of the domain structure by minimization of
the free energy density, containing two terms, magneto-static energy
and the surface energy of the domain walls
\cite{Kittel}-\cite{Chikazumi}. For observed CDs structures
\cite{Condon_Walstedt},\cite{Kramer1},\cite{Kramer2} the inequality
$D/L<<1$, where $L$ is the width of the sample, is fulfilled and
$\alpha \sim D/L<<1$.

The equation
\begin{equation} \label{eq:DPT}
a ( \mu_{0}H, T, T_{D} )= 1, \\
\end{equation}
defines the critical surface in three dimensions $\mu_{0}H-T-T_{D}$
. Above this surface the uniform diamagnetic phase exists, but below
it, the CD phase can appear in the part of every period of the dHvA
oscillations.

For proper calculation of the reduced amplitude of the dHvA
oscillations a and constructing the hysteresis loop, the correct
topology of Fermi surface has to be taken into account. In case of
Be the standard procedure of calculation of the amplitude of dHvA
oscillations \cite{Shoenberg}, based on series expansions near the
extreme cross-sections, requires considerable modification due to
negligible small curvature of extreme cross-sections of the
'cigar'-like part of the Fermi surface, contributed to the dHvA
oscillations \cite{Logoboy3}. The Fermi surface of Be consists of
the second zone monster ('coronet') and the third zone 'cigar'. It
is well-established that the dHvA oscillations originates from the
three maximum cross-sections of 'cigar' ('waist' and 'hips') which
are characterized by a very small curvature. In particular, the
small curvature of the cylinder like Fermi surface explains
relatively high amplitude of dHvA oscillations in Be at magnetic
field $\boldmath H$ applied parallel to the hexagonal axis, e.g.
$\boldmath H \parallel [0001]$, in comparing to Ag
\cite{Kramer1},\cite{Kramer3}, where the small amplitude of dHvA
oscillations is explained in the framework of spherical Fermi
surface \cite{Shoenberg}. In the framework of the model of slightly
corrugated cylinder like Fermi surface of Be \cite{Logoboy3}, the
reduced amplitude of the dHvA oscillations takes a form
\begin{equation} \label{eq:Reduced Amplitude}
a=\frac{4\kappa (\mathcal{A}_{0}\hbar)^{2}}{\pi^{3}m_{c}(\mu_{0}H)^{2}} |Q(\mu_{0}H)| R(\mu_{0}H,T,T_{D}), \\
\end{equation}
where $R(\mu_{0}H,T,T_{D})$ is a reduction factor \cite{Shoenberg},
\cite{Logoboy3}, which takes into account  the influence of the
temperature $T$ and impurities on the amplitude of the dHvA
oscillations.

The complex function
\begin{eqnarray} \label{eq:Q}
Q=Q_{1}+jQ_{2}=\mid Q\mid \exp{(j\psi)}, \qquad \qquad \nonumber \\
Q_{1}=3\beta \cos{\lambda}+2(1-2\beta) J_{0}(\lambda), \quad
Q_{2}=\beta \sin{\lambda}
\end{eqnarray}
where $\lambda=l^{2}\mathcal{A}_{0}\xi$, depends on applied magnetic
field through parameter $l^{2}=c\hbar/e\mu_{0}H$ and describes the
beatings in the amplitude of dHvA oscillations resulting from two
different extreme cross-section of Fermi surface of Be
\cite{Logoboy3}. Here, $\kappa=0.217 A^{o~-1}$ \cite{Shoenberg} is
the distance in reciprocal space between two extreme cross sections
of 'cigar', $A_{0}$ is average cross section area, $J_{0}(\lambda)$
is Bessel function of the first order and $\beta$ is the half-width
of the cylinder like extreme cross sections.

The Eqs.~(\ref{eq:DPT}),(\ref{eq:Reduced Amplitude}) define in
explicit form the critical temperature for diamagnetic phase
transition as a function of magnetic field $\mu_{0}H$ and Dingle
temperature $T_{D}$. It allows us to construct the phase diagrams
for Be and investigate the properties of the diamagnetic hysteresis.

\section{\label{sec:Results and Discussions}Results and Discussions}

\begin{figure}
  \includegraphics[width=0.4\textwidth]{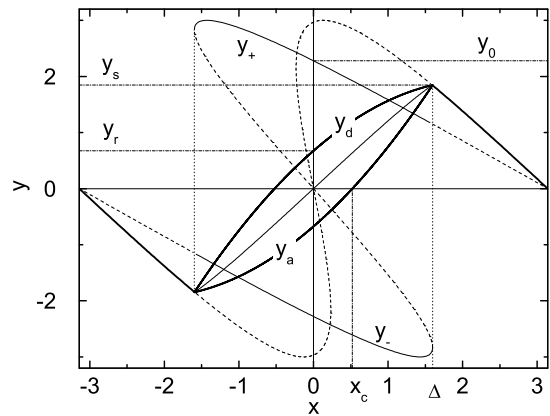}
\caption{ Diamagnetic magnetization loop $y=y(x;a)$ versus magnetic
field increment $x$ is plotted in one period of dHvA oscillations.
The local diamagnetic moments $y_{\pm}=y_{\pm}(x;a)$ reproduced from
\cite{Logoboy1} contribute to the hysteresis curve
$y_{a,d}=y_{a,d}(x;a)$. Three fundamental parameters of the
hysteresis loop, e.g. coercivity $x_{c}$, remnant magnetization
$y_{r}$ and saturation magnetization $y_{s}$ are shown. The dot
lines correspond to the range of existence of the hysteresis
($2\Delta$). The existence of the hysteresis is restricted by
$2\Delta \le 2\pi$. The dash lines show the states which are not
realized. In numerical calculations the value of $a=3$ and $n=0.52$
are used.} \label{Hysteresis}
\end{figure}

At the condition $a ( \mu_{0}H, T, T_{D} )>1$  the sample is divided
into domains with up and down magnetization. In the center of the
dHvA period $h=0$  the domains are characterized by the same width.
An external magnetic field removes the equivalence of the states,
the energy balance is altered and rearrangements of the domain
structure take place. For CD structure this process is realized not
only through the motion of domain walls, which results in
corresponding changes of domain volumes, and nucleation of the
domains, but also through magnetic field dependence of the local
magnetization in every adjacent domains \cite{Logoboy1}. To
understand the magnetization process in CD phase it is convenient to
define the approximate anhysteretic, or pinning-free, magnetization
\begin{equation} \label{eq:Anhysteretic Magnetization}
y^{(anhys)}=\frac {1}{2}(1-\frac{x}{\Delta})y_{-}+\frac {1}{2}(1+\frac{x}{\Delta})y_{+}, \\
\end{equation}
where the function $y_{\pm}=y_{\pm}(x)=-y_{\mp}(-x)$  is the
solution of the equation $y_{\pm}=a \sin {(x+y_{\pm})}$ at the range
$|x|\le \Delta \le k \cdot \Delta H/2$ in every period $\Delta H$ of
the dHvA oscillations and describes magnetic field dependence of the
local diamagnetic moments of the up ($+$) and down ($-$) domains
(Fig.~\ref{Hysteresis}). In Eq.~(\ref{eq:Anhysteretic
Magnetization}) we assumed that the volume fractions of the up and
down domains are linear functions of the increment of the reduced
magnetic field in every dHvA period. This assumption was justified
by NMR experiments in Ag \cite{Condon_Walstedt} and $\mu$SR
spectroscopy in Be \cite{Solt1}- \cite{Solt2}. Both experiments
reveal the domain structure by the appearance of a doublet
corresponding to two domain magnetic inductions with line
intensities, that show the linear dependence on the external
magnetic field of the volume fractions occupied by the neighboring
domains.

The range of existence of the CDs $2 \Delta$ is defined by
\cite{Logoboy1}
\begin{equation} \label{eq:Delta}
\Delta =\sqrt {a^{2}-1}- \cos^{-1}{ \frac {1}{a}} \ge 0, \quad a \ge 1. \\
\end{equation}
It follows from Eq.~(\ref{eq:Anhysteretic Magnetization}), that the
anhysteretic magnetization in every period of the dHvA oscillations
is a function of magnetic field $H$, temperature $T$ and Dingle
temperature $T_{D}$. The expression for $y^{(anhys)}$
Eq.~(\ref{eq:Anhysteretic Magnetization}) can be simplified for
relatively high reduced amplitude of the dHvA oscillations, e.g.
$a\gg 1$. In this case $y_{+} \approx y_{0}-(x/ \Delta)y_{0}$,
$y_{-} \approx y_{+}-2y_{0}$, where $y_{0}$ is the half-splitting in
the center of the period, defined explicitly form by the equation
$y_{0}/ \sin{y_{0}=a}$ \cite{Shoenberg}, and the
Eq.~(\ref{eq:Anhysteretic Magnetization}) is replaced by the
following one
\begin{equation} \label{eq:Approximation}
y^{(anhys)} \approx y_{0}(\frac {1}{\Delta}-\frac {1}{\pi})x, \quad |x| \le \Delta \le \pi. \\
\end{equation}
Although, the local diamagnetic moments $y_{\pm}=y_{\pm}(x_{ex})$
are characterized by average {\it negative} slope as the functions
of magnetic field increment $x_{ex}$ \cite{Logoboy1}, it follows
from Eq.~(\ref{eq:Approximation}), that the anhysteretic curve has a
{\it positive} slope in every period of the dHvA oscillations
(Fig.~\ref{Hysteresis}). We will see that the diamagnetic hysteresis
is also characterized by the positive slope: in the process of the
diamagnetic magnetization the volume fraction of the up-domains
increases on expense of the volume fraction of the down-domains with
increase of magnetic field increment $x_{ex}$ \cite{Shoenberg}.

The real diamagnetic magnetization process in CD phase is
characterized by hysteresis, or magnetization loop, which appears
periodically in applied magnetic field in every period of dHvA
oscillations, when the condition (\ref{eq:CDCondition}) is
fulfilled. According to experimental data [4] the diamagnetic
hysteresis is usually small, related to the domain wall displacement
and domain nucleation. Thus, it can be described in the framework of
the Lord Rayleigh model with two parameters: initial susceptibility
$\chi$ and Rayleigh constant $\eta$ (see, e.g. \cite{Chikazumi}):
\begin{equation} \label{eq:Rayleigh Model}
y_{a,d}= \pm y_{s}+\chi(x \mp \Delta) \pm \frac{1}{2} \eta (x \mp \Delta)^{2},\\
\end{equation}
where the upper (lower) sign corresponds to the ascending
(descending) magnetization in increasing (decreasing) magnetic field
(Fig.~\ref{Hysteresis}). The three fundamental parameters of the
hysteresis loop: the remnant magnetization $y_{r}=y_{r}(a)$, the
saturation magnetization $y_{s}$ and the coercivity $x_{c}=x_{a}$,
which is the reverse field to bring the magnetization to zero from
initial saturation, are characterized by strong dependence on the
applied magnetic field, temperature and Dingle temperature through
the reduced amplitude of the dHvA oscillations
$a=a(\mu_{0}H,T,T_{D})$. The saturation magnetization $y_{s}=y_{a}$
can be calculated numerically from the explicit equation
\begin{equation} \label{eq:Saturation Magnetization}
y_{s}=a \sin{[\Delta + (1-n)y_{s}]}.\\
\end{equation}
The Eq.~(\ref{eq:Rayleigh Model}) gives the following expressions
for the coercivity factor $x_{c}$ and remnant magnetization $y_{s}$:
\begin{eqnarray}
x_{c}=\frac{\delta\Delta^{2}}{1-\delta\Delta+\sqrt{(1-\delta\Delta)^{2}+(\delta\Delta)^{2}}}
\approx \delta\Delta^{2},\label{eq:Coercivity}\\
y_{r}=\frac{1}{2}\eta \Delta^{2}, \qquad \qquad
 \label{eq:Remnant}
\end{eqnarray}
where $\delta=\eta/\chi$. In calculation of
Eqs.~(\ref{eq:Coercivity}), (\ref{eq:Remnant}) we used $\delta
\Delta \ll 1$, which is justified by the observed narrow hysteresis
loop in Be \cite{Kramer2}. In this limit, coercivity factor $x_{c}$
and remnant magnetization $y_{r}$, being quadratic functions of the
range of CD existence $\Delta$ (\ref{eq:Delta}), have the same
temperature and magnetic field dependencies. The initial
susceptibility $\chi$ can be calculated from measured remnant
magnetization $y_{r}$ and coercivity factor $x_{c}$ according to the
equation $\chi=2y_{r}/x_{c}$. The hysteresis loop square which
characterizes the losses in the process of diamagnetic
magnetization, is defined as follows
\begin{equation} \label{eq:Losses}
S=\frac{4}{3}\eta \Delta^{3}.\\
\end{equation}
and also temperature and magnetic-field dependent. The loop square
$S$ Eq.~(\ref{eq:Losses}) reaches its maximum
$S_{max}=(4/3)\pi^{3}\eta$ at $a \gg 1$, inasmuch as the range of
the existence of the CDs $2\Delta$ in every period of the dHvA
oscillations is restricted by the period of the oscillations (
$2\pi$ in relative units). Analysis of the
Eqs.~(\ref{eq:Coercivity})-(\ref{eq:Losses}) with taking into
consideration Eq.~(\ref{eq:Delta}) shows that the parameters of the
hysteresis loop $x_{c}, y_{r}, S \to 0$ at the point of diamagnetic
phase transition, when $a \to 1$.

\begin{figure}
  \includegraphics[width=0.4\textwidth]{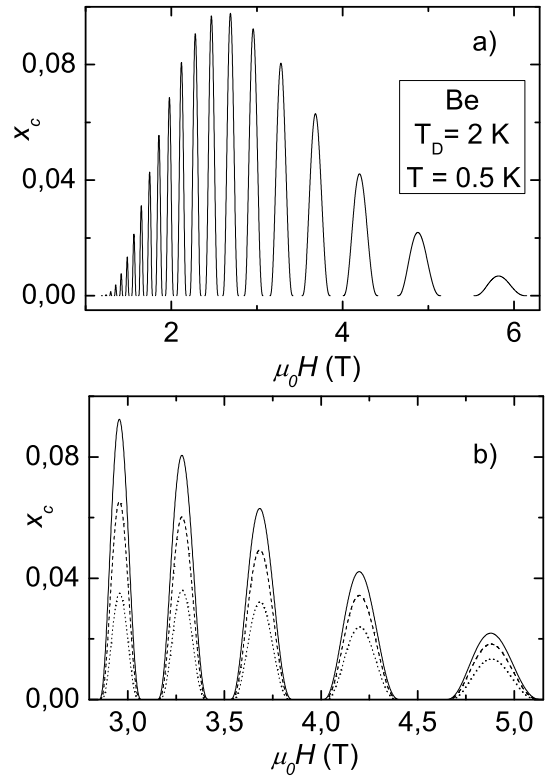}
\caption{\ Magnetic field dependence of the reduced coercivity
$x_{c}=x_{c}(\mu_{0}H)$ as a function of applied magnetic field at
$T=0.5K$ and $T_{D}=2K$ in the whole interval of CD existence in Be
(a) and at three different temperatures $T=0.5K$ (solid line),
$T=1K$ (broken line) and $T=1.5 K$ (dot line) (b) in the part of the
magnetic field interval. For calculations the value of $\delta=0.1$
is used.} \label{Magnetic Field Dependence}
\end{figure}

The magnetic field dependence of the coercivity
$x_{c}=x_{c}(\mu_{0}H)$ (\ref{eq:Coercivity}) at the Dingle
temperature $T_{D}=2K$ of the experiment \cite{Kramer2} and three
different temperatures $T$ from $0.5K$ to $1.5K$ is represented in
Fig.~(\ref{Magnetic Field Dependence}). To construct the plots we
used $\delta=0.1$, which is justified by comparison with the data
\cite{Kramer3}. The function $x_{c}=x_{c}(\mu_{0}H)$ is plotted at
the whole range of applied magnetic field $H$ from $1T$ till $6T$
(Fig.~\ref{Magnetic Field Dependence}(a)), where the non-uniform
diamagnetic phase exists at given Dingle temperature $T_{D}=2K$
\cite{Kramer3}. The function $x_{c}=x_{c}(\mu_{0}H)$ has asymmetric
multi-bell like shape, typical for the phase diagrams of Be
$\mu_{0}H-T-T_{D}$ \cite{Logoboy3}, \cite{Kramer3}. The maximums of
the function $x_{c}=x_{c}(\mu_{0}H)$ appear periodically on the
scale of inverse magnetic field with the period inversely
proportional to the discrepancy $\Delta F=F_{h}-F_{w}$ of the two
fundamental frequencies, $F_{h}$ and $F_{w}$, corresponding to two
extreme cross sections of cigar-like Fermi surface of Be:
\begin{equation} \label{eq:Period}
\Delta(\frac{1}{\mu_{0}H})=\frac{1}{\Delta F}.\\
\end{equation}
For $F_{h}=970.9T$ and $F_{w}=942.2T$
\cite{Shoenberg},\cite{Kramer2} we obtain $\Delta F=28.7T$. The
increase of the temperature $T$ results in decrease of the maximums
of the function $x_{c}=x_{c}(\mu_{0}H)$ till its disappearing, when
the point of phase transition is reached ($a=1$). The position of
the maximums of the function $x_{c}=x_{c}(\mu_{0}H)$ does not depend
on temperature in accordance with the phase diagrams for Be
\cite{Logoboy3}. The periodicity of the coercivity on reciprocal
magnetic field is illustrated in Fig.~\ref{Inverse Magnetic
Field}(a), where the coercive force $X_{c}=x_{c}/k=X_{c}(\mu_{0}H)$
measured in Gauss is plotted as a function of $(\mu_{0}H)^{-1}$ at
$T=1.3K$ and $T_{D}=1.8K$. It follows from Fig.~\ref{Inverse
Magnetic Field}(a), that the preferable interval of applied magnetic
field for measuring the coercive force is $2-4T$ , where the
function $X_{c}$ can reach maximum values from $1G$ to $2G$ at given
values of $T=1.3K$ and $T_{D}=1.8K$.

\begin{figure}[b]
  \includegraphics[width=0.4\textwidth]{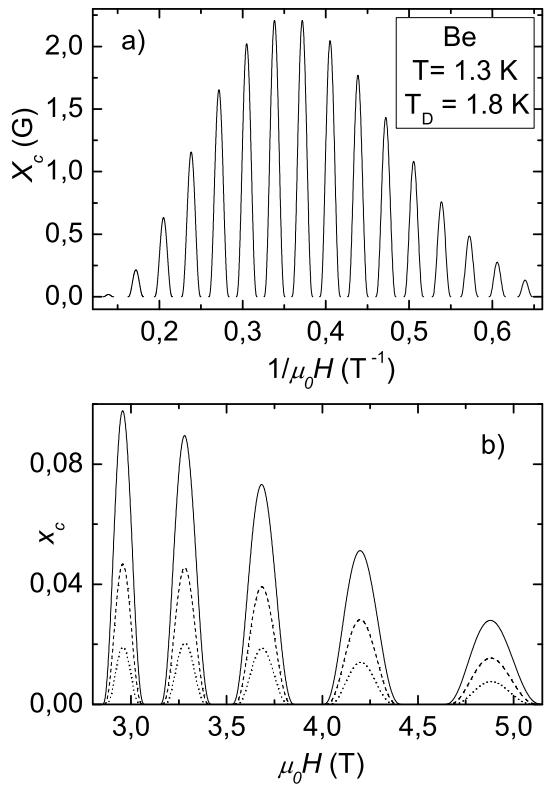}
\caption{(a) Coercive force $X_{c}=x_{c}/k=X_{c}(\mu_{0}H)$ as a
function of inverse magnetic field at $T=1.3K$ and $T_{D}=1.8K$, and
(b) reduced coercivity $x_{c}=x_{c}(\mu_{0}H)$ at $T=1.3K$ and three
different Dingle temperatures: $T_{D}=1.8$ (solid line), $T_{D}=2K$
(broken line) and $T_{D}=2.2K$ (dot line).} \label{Inverse Magnetic
Field}
\end{figure}

The increase of the Dingle temperature $T_{D}=\hbar /2\pi k_{B}\tau$
which is inversely proportional to the scattering lifetime $\tau$ of
the conduction electrons, leads to the reduction of the amplitude of
the dHvA oscillations, resulting in the phase transition into
uniform diamagnetic phase. Thus, the range of the existence of the
CD phase $\Delta$ Eq.~(\ref{eq:Delta}) decreases ($\Delta \to 0$,
when $a \to 1$). Assuming that this contribution to the coercive
force $x_{c}$ Eq.~(\ref{eq:Coercivity}) is dominant and neglecting
an unessential change of the parameter $\delta=\chi/\eta$, one can
come to the bells-like magnetic field dependencies of $x_{c}$, as
illustrated in Fig.~\ref{Inverse Magnetic Field}.

\begin{figure}
  \includegraphics[width=0.4\textwidth]{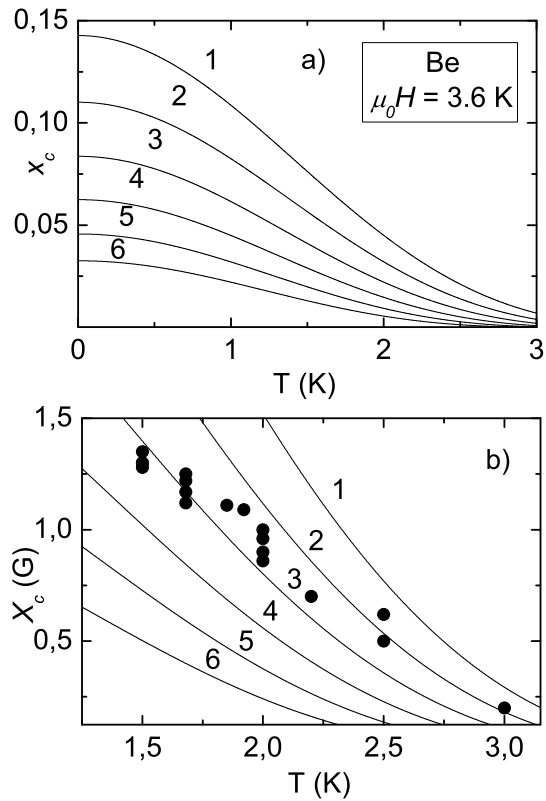}
\caption{(a) Temperature dependence of the reduced coercivity factor
$x_{c}=x_{c}(T)$ and (b) coercive force $X_{c}=x_{c}/k=X_{c}(T)$ at
$\mu_{0}H=3.6T$ and six different Dingle temperatures $T_{D}:
1~-~1.8K, 2~-~1.9K, 3~-~2K, 4~-~2.1K, 5~-~2.2K, 6~-~2.3K$. Close
circles are referred from R.Kramer {\it et al
}\cite{Kramer3}.}\label{Temperature Dependence}
\end{figure}

The temperature dependencies of the reduced coercivity factor
$x_{c}=x_{c}(T)$ and measured coercive force $X_{c}=X_{c}(T)$ at
$\mu_{0}H=3.6T$ and different Dingle temperatures are shown in
Fig.~\ref{Temperature Dependence}. The circles in
Fig.~\ref{Temperature Dependence}(b) represent the results of
measurement of the coercive force in Be \cite{Kramer3}. The
theoretical curve $X_{c}=X_{c}(T)$ plotted at$\mu_{0}H=3.6T$ and
$T_{D}\approx 1.8K$ fits the experimental points. The reported
Dingle temperature in \cite{Kramer3} is $T_{D}=2K$. There is a good
agreement between predicted temperature dependence of the coercive
force and data \cite{Kramer3}.

The observation of the hysteresis loop in Be \cite{Kramer1} provides
the possibility of the verification of the developed theory of the
diamagnetic hysteresis and evaluate parameters of the loop. The
periodical irreversible behavior of the magnetization process in the
rod-like sample $8\times2\times1mm^{3}$ of Be is demonstrated at
$T=1.3K$ in the whole range of applied magnetic field $\mu_{0}H
\approx 2-6T$, where the CD phase exists. The Dingle temperature is
$T_{D}=2K$. The diamagnetic phase diagrams estimated from
Eq.~(\ref{eq:DPT}) with taking into account the Eq.~(\ref{eq:Reduced
Amplitude}) for the reduced amplitude of oscillations $a$, obtained
in the framework of the slightly corrugated Fermi surface of Be
\cite{Logoboy3}, are consistent with the data \cite{Kramer2},
\cite{Kramer3}. At $\mu_{0}H=3.6T$ the calculated period of the dHvA
oscillations $\Delta H=2\pi/k=(\mu_{0}H)^{2}/F_{0} \approx 138G$
($F_{0}=955T$ is average area of the extreme cross-sections of the
Fermi surface sheet, relevant to dHvA oscillations \cite{Shoenberg})
is very close to the observed period of the dHvA oscillations
$\Delta H \approx 134-138G$ \cite{Kramer2}. According to the data
\cite{Kramer2}, the coercive force $X_{c}\approx 1.2G$, the remnant
magnetization $Y_{r}=y_{r}/k\approx 1.2G$, the saturation
magnetization $Y_{s}$ and the range of existence of the non-uniform
phase $2X_{\Delta}$ can be evaluated as $Y_{s}\approx 30G$ and
$2X_{\Delta}\approx 34G$ consequently. Thus, the experimental value
of the reduced parameter $\Delta^{(exp)}=X_{\Delta}k\approx 0.8$ can
be compared with the corresponding theoretical value of $\Delta$
Eq.~(\ref{eq:Delta}) in the following way. At $T=1.3K, T_{D}=2K$ and
$\mu_{0}H=3.6T$, the reduced amplitude of the dHvA oscillations
according to the Eq.~(\ref{eq:Reduced Amplitude}) is $a\approx 2$,
giving the value of $\Delta\approx0.7$ in accordance with the
experimental value $\Delta^{(exp)}=X_{\Delta}k\approx 0.8$
\cite{Kramer2}. According to this calculation, at the conditions of
the experiment \cite{Kramer2}, CD phase occupies $\approx 0.25$ part
of the period. The data \cite{Kramer2} allow us to evaluate the
parameters of the Rayleigh hysteresis loop, $\chi$ and $\eta$.
First, we calculate $y_{r}=kY_{r}\approx0.053$, that gives
$\eta=2y_{r}/\Delta^{2}\approx0.18$, $\chi=2y_{r}/x_{c}\approx2$ and
$\delta=0.09\approx0.1$. Then, we can compare the experimental and
theoretical values for saturation magnetization. According to the
data \cite{Kramer2} $y^{(exp)}_{s}=kY_{s}\approx1.4$. The
theoretical value of the saturation magnetization $y_{s}$ can be
calculated from Eq.~(\ref{eq:Saturation Magnetization}). Using the
value of the demagnetizing coefficient $n\approx0.05$, calculated
for prolate ellipsoid \cite{Osborn}, $a\approx2$ and
$\Delta\approx0.7$ one can obtain $y_{s}\approx1.6$ in accordance
with the experimental value of $y^{exp}_{s}=1.4$.

\section{Conclusions}

Magnetic interaction of strongly correlated electron gas results in
non-linear relationship between magnetization and magnetic field,
which can give rise an effect of hysteresis in magnetization curve
due to existence of metastable states for local diamagnetic moments
in each period of dHvA oscillations. There are three main
contributions into the process of magnetization of normal metals in
the condition of strong dHvA effect: the domain wall displacement,
the nucleation of domains and the dependence of local diamagnetic
moments on external magnetic field \cite{Logoboy1}.

We developed the model of diamagnetic hysteresis in CD phase in the
framework of Rayleigh hysteresis loop. Rayleigh law for the
magnetization process of spin origin is generally believed to be
caused only by the wall displacement and in relatively small
magnetic field. There are two circumstances that can justify the
applicability of the Rayleigh model for magnetization process of
strongly correlated electron gas. The first, although for
observation of the dHvA effect the high magnetic field in the scale
of $1-10T$ is need, the interplay between uniform and non-uniform
phases realized at the fine magnetic field scale of the magnetic
field increment $|h|\le\Delta H=(\mu_{0}H)^{2}/F$, e.g. $\sim mT$,
in every period of the dHvA oscillations. And, the second, the
magnetization process for electron gas in the conditions of dHvA
effect is dominated by domain wall motion which is similar to the
magnetization process of soft thin magnetic films in the physics of
magnetism of spin origin. There is a good agreement between the
theoretical results obtained in the framework of the developed model
of diamagnetic hysteresis and recent experimental results on
observation of the hysteresis in Be \cite{Kramer2}. The diamagnetic
hysteresis is usually small, appears periodically in every period of
dHvA oscillations with changing the value of applied magnetic field,
and is described in the framework of the Rayleigh model with two
fundamental parameters: initial susceptibility $\chi$ and Rayleigh
constant $\eta$. We calculate the parameters of the diamagnetic
hysteresis from  the data \cite{Kramer2}. The zero field saturation
(remnant) magnetization $y_{c}$, coercivity $x_{c}$ and hysteresis
loop square $S$ are characterized by strong dependence on
temperature, magnetic field and impurity of the sample through the
reduced amplitude of the dHvA oscillations $a(\mu_{0}H,T,T_{D})$.
The irreversible changes in magnetization are the result of
existence of metastable states due to defects of crystal lattice,
intrinsic stress and deformation. Generally, the diamagnetic
hysteresis occurs because the domain boundaries do not return to
their original positions when the external magnetic field approaches
to its value corresponding to the centre of period of dHvA
oscillations. At the initial permeability range the domain walls are
displaced reversibly from their stable positions which essentially
determined by the homogeneity of materials. Beyond the initial
permeability range, the intensity of magnetization increases and the
process is irreversible. The irreversible magnetization range is
attained by irreversible displacement of domain walls. Although the
rotation processes, that are intrinsic property of magnetization
loop in magnetic materials of spin origin, can be neglected in
rearrangement of CD structure, the microscopic theory of diamagnetic
loop has to include into consideration the possible bending of the
domain wall due to magnetization current densities localized in the
domain wall close to the sample surface \cite{Logoboy2}. The
irreversibility of the magnetization process is believed to be the
intrinsic property of the CD structure for all normal metals. The
present model of diamagnetic hysteresis can be applied for
investigation of temperature and magnetic field dependencies of
diamagnetic magnetization not only in Be, but in other systems,
where the CDs were observed. We hope that our theoretical results
will motivate the experimentalists to carry out further experiments
in investigation such an exotic phenomenon as diamagnetic
hysteresis.

\begin{acknowledgments}
We are indebted to V. Egorov and I. Sheikin for illuminating
discussions. We are also grateful to R. Kramer for providing us with
his experimental results before publication.
\end{acknowledgments}

\end{document}